\documentclass[fleqn,usenatbib]{mnras}

\usepackage{newtxtext,newtxmath}

\usepackage[T1]{fontenc}

\DeclareRobustCommand{\VAN}[3]{#2}
\let\VANthebibliography\thebibliography
\def\thebibliography{\DeclareRobustCommand{\VAN}[3]{##3}\VANthebibliography}

\usepackage{graphicx}	
\usepackage{amsmath}	
\usepackage{tablefootnote}

\title[Young HMXBs in SNRs]{Origin of young accreting neutron stars in high-mass X-ray binaries in supernova remnants}

\author[A. D. Khokhriakova, S. B. Popov]{
A. D. Khokhriakova,$^{1,2,3}$\thanks{E-mail: alenahohryakova@yandex.ru}
S. B. Popov,$^{1,2}$
\\
$^{1}$Department of Physics, Lomonosov Moscow State University, Moscow, 119991 Russia\\
$^{2}$Sternberg Astronomical Institute, Lomonosov Moscow State University, Moscow, 119234 Russia\\
$^{3}$``Basis'' foundation fellow\\
}

\date{Accepted XXX. Received YYY; in original form ZZZ}

\pubyear{2021}

\begin{document}
\label{firstpage}
\pagerange{\pageref{firstpage}--\pageref{lastpage}}
\maketitle

\begin{abstract}
Recently, several accreting neutron stars (NSs) in X-ray binary systems inside supernova remnants have been discovered. They represent a puzzle for the standard magneto-rotational evolution of NSs, as their ages ($\lesssim 10^5$ years) are much shorter than the expected duration of Ejector and Propeller stages preceding the onset of wind accretion. 
To explain appearance of such systems, we consider rotational evolution of NSs with early fallback accretion and asymmetry in forward/backward transitions between Ejector and Propeller stages (so-called hysteresis effect proposed by V. Shvartsman in 1970).
It is shown that  after a successful fallback episode with certain realistic values of the initial spin period, stellar wind properties, and magnetic field, a young NS may not enter the Ejector stage during its evolution which results in a relatively rapid initiation of accretion within the lifetime of a supernova remnant. For a standard magnetic field $\sim 10^{12}$~G and initial spin period $\sim 0.1$~--~0.2~s accretion rate $\gtrsim 10^{14}$~--~$10^{15}$~g~s$^{-1}$ is enough to avoid the Ejector stage. 
\end{abstract}
\begin{keywords}
stars:neutron --  X-rays: binaries -- stars:supernovae 
\end{keywords}



\section{Introduction}

Evolution of neutron stars (NSs) is of great interest for astrophysics and fundamental physics. Due to sophisticated interplay between initial distributions of the main parameters, 
their evolution, and properties of surrounding medium detailed understanding of many stages of NSs life is still lacking. In order to fill gaps in our theoretical description of NS evolution we need more observations of these objects at different ages, including young sources in non-trivial conditions.

In recent years, accreting NSs in X-ray binary systems located in supernova remnants (SNRs) have been discovered. They are specifically interesting objects  allowing to probe early evolution of accreting NSs as SNRs have a typical lifetime $\lesssim10^5$~yrs. Given short time scale, such sources are quite rare. At the moment only a few are known (\citealt{2021arXiv210709325X}, see Table~\ref{tab:HMXB}). 

SXP 1062 and SXP 1323 both have Be stars as companions and are located in the Small Magellanic Cloud. { An interesting feature of SXP 1062 is its large negative period derivative $\dot P \approx -0.24  \text{ s day}^{-1}$ \citep{2012A&A...537L...1H, 2018MNRAS.475.2809G}. 
As of 2019, its period was 979.5 s \citep{2020A&A...637A..33T}.
SXP 1323 also has a negative $\dot P \approx -0.08 $ s day$^{-1}$ \citep{2021arXiv210705589M}.} 

DEM L241 is reliably a high-mass X-ray binary (HMXB). However, the nature of the compact object in DEM L241 is uncertain. The optical counterpart of DEM L241 is an O5III(f) star with mass of about 40 $M_\odot$ \citep{2012ApJ...759..123S}. Correspondingly, the progenitor of the compact object should have been massive, which makes formation of an NS less probable.

On the contrary, Circinus X-1 definitely has an NS as an accretor, but it is unclear, whether it is a HMXB or a low-mass X-ray binary (LMXB) \citep{2010ApJ...719L..84L}. It is a highly variable X-ray source and the distance to this object is not very certain \citep{2015ApJ...806..265H}. These two factors result in an uncertainty in its luminosity. 
Observations with the Very Large Telescope suggest that, if the spectrum belongs to the companion star, it has to be a supergiant \citep{2007MNRAS.374..999J}. 

Among known accreting X-ray binaries in SNRs the recently discovered by \cite{2019MNRAS.490.5494M} HMXB {  XMMU J051342.6-672412} in the supernova remnant MCSNR J0513-6724 is of particular interest { (hereafter we use for this source notation LXP~4.4 introduced in \citealt{2021MNRAS.504..326M})}. 
 The age estimate for this system ($< 6 \times 10^3$ years) makes it one of the youngest  known sources of this type and challenges researchers to explain how  a NS can reach the accretion stage in such a short time. 

\subsection{Standard magneto-rotational evolution of a neutron star} 

Let us resume the standard scenario of NS evolution following systematization presented e.g., by \citet{1992ans..book.....L}. 
By default, a NS is born at the stage of Ejector. 
This is the stage when surrounding material cannot reach surface or even magnetosphere of the NS mainly due to a powerful relativistic wind generated and accelerated by the compact object.
The duration of this stage can be quite long:
\begin{equation}
    t_\mathrm{E} = \frac{3 I c^3}{8 \upi^2 
    \mu^2}(p_\mathrm{E}^2 - p_0^2) \approx 1.1 \times 10^7 \text{ yr } \bigg(\frac{B}{10^{12} \text{ G}}\bigg)^{-2}, 
    \label{eq:t_ej}
\end{equation}
where 
$I = 10^{45}$ g cm$^2$ is moment of inertia of the NS,
$c$ is the speed of light. 
Two values of the spin period $p_0$ and $p_\mathrm{E}$ correspond to the initial value and to the final value at the Ejector stage, correspondingly.
Finally, $\mu = B R_{\mathrm{x}}^3$ is a magnetic dipole moment, here $B$ is the magnetic field on the equator of the NS, $R_{\mathrm{x}} = 10$ km is the radius of the NS. 

Eq. (\ref{eq:t_ej}) is derived under the assumption that the magnetic field is constant, and the loss of rotational moment is described by the standard magneto-dipole formula:
\begin{equation}
    \frac{1}{2} \frac{\mathrm{d}(I\omega^2)}{\mathrm{d}t} = -\frac{2}{3} \frac{\mu^2 \omega^4}{c^3} \sin^2 \chi,
    \label{magrot}
\end{equation}
where $\chi$ is an angle between rotational and magnetic axis, which is assumed to be constant and equal to $\chi=\arcsin (1/\sqrt{2})$ everywhere below. We use this formula for calculation of spin down at the Ejector phase.

After the Ejector stage the star can proceed to the Propeller stage, and then to Accretor or Georotator (see below).
Transition from stage to stage can be specified in terms of pressure balance. Pressure related to the external, i.e. infalling, matter (gravitationally captured or not) has to be equalized by magnetospheric pressure or ram pressure of the wind or/and radiation. 
However, from practical point of view it is useful to reformulate transition conditions in terms of critical periods and radii. 

We start with gravitational capture radius, as definitions of several other critical values depend on this quantity:
\begin{equation}
    R_\mathrm{G} = \frac{2 G M_\mathrm{x}}{v^2} \approx 3.7 \times 10^{10} \bigg(\frac{v}{10^8 \text{ cm s$^{-1}$}}\bigg)^{-2} \text{cm} ,
\end{equation}
where $M_\mathrm{x}$ is mass of the compact object. In all our calculations and estimates we use $M_\mathrm{x} = 1.4 M_{\sun}$. Parameter $v$ is velocity of the surrounding medium relative to the NS. Depending on the exact situation this velocity can be related to motion of an isolated NS in the interstellar medium, or/and to the sound speed in the medium, or/and to the wind velocity of a donor in a binary system, or/and to the orbital velocity, or to  a combination of some of these values. { We mainly present results for $v=10^8$~cm~s$^{-1}$, as this is a typical value for massive stars in close binary systems.}

The magnetospheric radius is assumed to be one half (see, e.g. \citealt{1996ApJ...465L.111W}, \citealt{2018A&A...610A..46C}) of the Alfven radius:
\begin{equation}
    R_{\mathrm{m}} = 0.5 R_\mathrm{A} = 
    \begin{cases}
    0.5 \times \bigg(\frac{2 \mu^2 G^2 M_\mathrm{x}^2}{\dot{M} v^5}\bigg)^{1/6},&\text{if $R_\mathrm{A}>R_\mathrm{G}$;}\\
    0.5 \times \bigg(\frac{\mu^2}{2 \dot{M} \sqrt{2 G M_\mathrm{x}}}\bigg)^{2/7},&\text{if $R_\mathrm{A}<R_\mathrm{G}$.}
    \end{cases}
\end{equation}
In these equations $\dot M$ is the accretion rate. Note, that it can be defined even if accretion on the surface is not happening: $\dot M = \upi R_\mathrm{G}^2 \rho v$. It is just a combination of surrounding matter density and its velocity relative to the compact object. Thus, $\dot M$ just characterizes properties of the external medium in its interaction with the NS and its radiation.

A very important quantity for our discussion is the so-called
Shvartsman radius \citep{1970SvA....14..527S}. It is defined by equality of the pressure of relativistic wind from a NS and external pressure:
\begin{equation}
\begin{gathered}
    R_\mathrm{Sh} = \bigg(\frac{8 
    \mu^2 (G M_\mathrm{x})^2 \omega^4}{3 \dot{M} v^5 c^4}\bigg)^{1/2} \approx \\ \approx 
    1.3 \times 10^{10}\,  p^{-2} \bigg(\frac{B}{10^{12} \text{ G}}\bigg) \bigg(\frac{\dot M}{10^{14} \text{ g s}^{-1}}\bigg)^{-1/2} \bigg(\frac{v}{10^8 \text{ cm s}^{-1}}\bigg)^{-5/2}\, \mathrm{ cm}. 
\end{gathered}
\end{equation}
Transition to/from the Ejector stage depends on this parameter. If $R_\mathrm{Sh}>R_\mathrm{G}$, then a stable cavern can form around an NS.  

Another critical distance important for the Ejector stage is the well-known light cylinder radius:
\begin{equation}
    R_\mathrm{l} = \frac{c p}{2 \upi} \approx  4.8 \times 10^9 \, p \, \text{ cm}. 
\end{equation}

In our code and graphs we also use a general term ``stopping radius'', $R_\mathrm{stop}$, which is equal to the distance where a pressure balance is reached. At the Acrretor and Propeller stages the stopping radius is situated inside the light cylinder: $R_\mathrm{stop} = R_\mathrm{m}$. At the Ejector stage the stopping radius is outside the light cylinder and is larger than the gravitational capture radius: $R_\mathrm{stop} = R_\mathrm{Sh}$.

Finally, Accretor/Propeller transition depends on the
corotation radius:
\begin{equation}
    R_\mathrm{c} = \bigg(\frac{G M_\mathrm{x} p^2}{4 \upi^2}\bigg)^{1/3} \approx 1.7 \times 10^8 \, p^{2/3} \, \text{ cm}.
\end{equation}

In the standard scenario the Ejector stage is terminated when due to spin-down and corresponding decrease of the relativistic wind power $R_\mathrm{Sh}$ diminishes down to $R_\mathrm{G}$. Then external matter becomes gravitationally captured and its pressure grows toward the NS. As a result, matter can penetrate inside the light cylinder which ceases particle production and acceleration in the magnetosphere. 
After the transition normally $R_\mathrm{m}>R_\mathrm{c}$, so we have a Propeller.  Rapid spin-down result in growing of $R_\mathrm{c}$. Finally, the NS enters the stage of accretion (or, if $R_\mathrm{m}>R_\mathrm{G}$ --- the Georotator stage). 

Spin-down at the Ejector stage was specified above in eq.~(\ref{magrot}). 
At propeller phase spin-down is very uncertain. The only clear statement is that it is more rapid than at the Ejector. In this study we assume the following approach \citep{1975SvAL....1..223S}: 
\begin{equation}
    \frac{\mathrm{d}p}{\mathrm{d}t} = \frac{p^2}{2 \upi I} K_\mathrm{sd}^\mathrm{(P)},
    \label{eq:p_dot_prop}
\end{equation}
where
\begin{equation}
    K_\mathrm{sd}^\mathrm{(P)} = \dot{M} \frac{2 \upi}{p} R_\mathrm{m}^2,
\end{equation}
is the spin-down (breaking)  moment at the Propeller stage.

At the stage of accretion both --- spin-up and spin-down, --- momenta coexist:
\begin{equation}
    \frac{dp}{dt} = \frac{p^2}{2 \upi I} (K_{\text{sd}}^\text{A} - K_{\text{su}}^\text{A}).
    \label{eq:p_dot_acc}
\end{equation}

Spin-down does not include properties of the external medium:
\begin{equation}
    K_{\text{sd}}^\text{(A)} = k_\mathrm{sd}\frac{ \mu^2}{R_c^3}.
    \label{eq:K_sd_A}
\end{equation}
Here $k_\mathrm{sd}$ is a dimensionless coefficient, which can depend on parameters of the problem.

Oppositely to $K_{\text{sd}}$, spin-up (we write for the case of wind accretion) depends on the accretion rate and orbital frequency $\Omega$:
\begin{equation}
    K_{\text{su}}^\text{(A)} = \dot{M} \eta_\mathrm{t} \Omega R_\mathrm{G}^2,
    \label{eq:K_su_A}
\end{equation}
where $\eta_t$ is a dimensionless coefficient, { we assume $\eta_\mathrm{t}=0.25$ \citep{1992ans..book.....L}.} 
Typically, it is assumed that at the Accretor stage spin period relaxes to the so-called equilibrium value, which is determined by equality between spin-up and spin-down moments. This assumption is often used to estimate the magnetic field as $\dot M$ can be directly derived from the observed flux and known distance to the source.
{The equilibrium period can be obtained by equating expressions (\ref{eq:K_sd_A}) and (\ref{eq:K_su_A}):
\begin{equation}
    p_{\text{eq}} = \bigg(\frac{2 \pi k_\text{t} \mu^2}{\eta_\text{t} G M_\mathrm{x} \dot M R_G^2} P_\text{orb} \bigg)^{1/2}.
\end{equation}
}

\begin{table*}
	\caption{Known HMXBs in SNRs}
	\label{tab:HMXB}
		\begin{flushleft}
	\begin{tabular}{lccccc} 
		\hline
		Name & Spin period, & Luminosity, & Age,  & Orbital & Refs.\\
		& & & & period, \\
		     & s    & erg s$^{-1}$ & yrs &  days\\
		\hline
		SXP 1062 & 1062 & $6.9 \times 10^{35}$ & $(2-4)\times 10^{4}$ & $656$ & \citet{2012MNRAS.420L..13H} \\ 
		& & & & & \citet{2021MNRAS.503.3856G} \\
		SXP 1323 & 1323 & $10^{35}$ -- few $\times 10^{36}$ & $\sim 4 \times 10^{4}$ & 26.2 & \citet{2019MNRAS.485L...6G}\\
		DEM L241 & --- & $2 \times 10^{35}$ & - & $10.3$ & \citet{2012ApJ...759..123S},\\
		& & & & & \citet{2016ApJ...829..105C}\\
		LXP 4.4 & 4.4 & $7 \times 10^{33}$ & $< 6 \times 10^3$  & 2.2 & \citet{2019MNRAS.490.5494M}\\
		Circinus X-1$^{*}$ & --- & $\sim 10^{35**}$ & $< 4.6 \times 10^3$ & 16.5 & \citet{2013ApJ...779..171H} \\
		XMMU$^{***}$  & 570 & $9 \times 10^{34}$ & $(43-63)\times 10^{3}$ & 40.2$^{****}$ & \citet{2021MNRAS.504..326M}\\
		J050722.1-684758 & & & \\
		\hline
	\end{tabular}
	\\
	$^{*}$ Might be an LMXB.\\
	$^**$Luminosity corresponds to a dim state of the source, see \cite{2020ApJ...891..150S}. \\
	Circinus X-1 has a highly variable X-ray luminosity that can
	reach $> 10^{38}$ erg s$^{-1}$.\\
	$^{***}$ Physical association with SNR is still unclear. \\
	$^{****}$ \citet{2018MNRAS.475.3253V} suggested orbital period of 5.27 days.
	\end{flushleft}
\end{table*}

Duration of Ejector and Propeller stages is determined by how rapidly the NS spins-down. In usual conditions (including standard magnetic fields $\sim 10^{12}$~G and short initial periods) it can take million years.  
HMXBs in SNRs are young objects which do not easily fit the standard scenario.
The recently discovered system LXP~4.4 is especially challenging. In the first place, its age is less than 6000 years. In some cases (see e.g. \citet{2012MNRAS.421L.127P} on SXP~1062) rapid transition can be explained by large initial magnetic field. But in the case of LXP~4.4 this is not an option as the spin period of accretor is small as well as the luminosity, and so the condition $R_\mathrm{m}<R_\mathrm{c}$ cannot be reached for large fields. \cite{2019MNRAS.490.5494M} provide an estimate $B\sim 3 \times 10^{11}$~G basing on $p=4.4$~s and assuming $R_\mathrm{m}=R_\mathrm{c}$.


Accounting for the fact that in the standard scenario a young NS is unlikely to be an Accretor, \cite{2020MNRAS.494...44H}  propose that the observed X-ray luminosity might be mainly  due to thermal emission from a young cooling magnetized NS with a probable small additional contribution from
magnetospheric accretion in the Propeller phase.

Another obvious explanation is that the NS in LXP~4.4 was born with a long spin period, { close to the present-day value}. Thus, it quickly (or even immediately) appeared at the stage of accretor, where later its period  relaxes to the equilibrium value.

We consider here another idea involving initial fallback stage leading to absence of the Ejector phase. In the next section we describe basics of our hypothesis and discuss expected fallback parameters. In Sec. 3 we present results of our calculations. In the final section we discuss uncertainties together with possible alternatives and present our conclusions.

\section{Evolution of a NS in a binary accounting for hysteresis effect and fallback}

\subsection{Hysteresis}

Transitions between stages along a track of the magnetorotational evolution of an NS are determined by pressure balance at critical radii. However, this equality can be reached at different radii for forward and backwards directions.\footnote{We call ``forward'' transitions in the sequence Ejector~$\rightarrow$~Propeller~$\rightarrow$~Accretor.} Thus, the term ``hysteresis'' is used to describe this situation.  In particular, we are interested in asymmetry at Ejector-Propeller transition, first noticed by \cite{1970SvA....14..527S}. We refer it throughout the paper simply as ``hysteresis''. 

In the phase of ejection the NS slows down. It results in decrease of the wind power and Shvartsman radius. Gradually, $R_\mathrm{Sh}$ approaches the gravitational capture radius. 
Forward transition Ejector~$\rightarrow$~Propeller happens when external pressure equalizes the relativistic wind pressure at $R_\mathrm{G}$. 
Pressure of gravitationally captured cold matter grows towards the NS with distance as $r^{-5/2}$, i.e. faster than the wind pressure ($\sim r^{-2}$). So, as described in Sec 1, particle generation and acceleration are stopped. 

 Transition in the opposite direction (Propeller$\rightarrow$Ejector) proceeds in a different way. When the propeller phase is on and accretion rate decreases (which happen e.g. under the fallback conditions), we have $R_\mathrm{c}<R_\mathrm{m}<R_\mathrm{l}$ and the magnetospheric radius is growing gradually approaching the light cylinder. 
However, at the light cylinder the pressure of external gravitationally captured matter is large. To overcome it is necessary to have much larger power than in the case of the equilibrium at $R_\mathrm{G}$. { I.e., the transition is characterized by a shorter critical period.}

This asymmetry can be illustrated by values of critical period at forward and backward transitions.
For the usual case when $R_\mathrm{G}$ > $R_\mathrm{l}$ we obtain for E $\rightarrow$ P transition { from the condition $R_\mathrm{Sh}=R_\mathrm{G}$}:
\begin{equation}
\begin{gathered}
    p_\mathrm{E\rightarrow P} = \frac{2 \upi}{c} \bigg(\frac{2 k_t 
    \mu^2}{3 v \dot{M}}\bigg)^{1/4} = \\ =
    0.6 \text{ } \bigg(\frac{B}{10^{12} \text{ G}}\bigg)^{1/2} \bigg(\frac{\dot M}{10^{14} \text{ g s}^{-1}}\bigg)^{-1/4} \bigg(\frac{v}{10^8 \text{ cm s}^{-1}}\bigg)^{-1/4}\, \mathrm{ s} ; 
\end{gathered}
\end{equation}
and for the reverse one (P $\rightarrow$ E) { from equating relativistic wind pressure to the pressure of the ambient medium at $R_\mathrm{l}$} we obtain:
\begin{equation}
\begin{gathered}
    p_\mathrm{P \rightarrow E} = \frac{2 \upi}{c} \bigg(\frac{\mu^4}{8 G M_\mathrm{x} \dot{M}^2}\bigg)^{1/7} = \\ = 0.19 \text{ } \bigg(\frac{B}{10^{12} \text{ G}}\bigg)^{4/7} \bigg(\frac{\dot M}{10^{14} \text{ g s}^{-1}}\bigg)^{-2/7} \mathrm{ s} .
\end{gathered}
\end{equation}
{Note, that the last equation formally is obtained from $R_\mathrm{A}=R_\mathrm{l}$.}

{ Note different dependencies of two critical periods on the velocity. Difference between $p_\mathrm{E\rightarrow P}$ and $p_\mathrm{P \rightarrow E}$ is smaller for larger velocities. 
However, the existence of hysteresis is guaranteed while the condition $R_\mathrm{G} > R_\mathrm{l}$ is met.}

\citet{1970SvA....14..527S} applied the hysteresis to a situation when an NS, initially at the Ejector stage, experiences enhanced rate of accretion for some period of time. When this episode is over, the NS can avoid turning back to Ejector staying at an Accretor or Propeller stage. 
Here we apply the same effect to the situation when an NS after birth experiences a period of intense fallback { (thus, passing through the stage of accretion)}, and then while $\dot M$ is decreasing switches to Propeller, but can avoid becoming an Ejector due to hysteresis. Thus, period of time prior to the initiation of accretion can be much shorten in such case. 


\subsection{Fallback}

After formation of an NS, some amount of the progenitor star material expelled in the supernova explosion can fall back onto the compact object (see e.g. an essential paper \citealt{1989ApJ...346..847C} and references to early studies therein). In case of NSs, this amount can be up to few$\times0.1\, M_\odot$, however, more typical values might be about $\lesssim 10^{-3}\, M_\odot$. 
Fallback can significantly influence various properties of NSs, including spin period and magnetic field (see recent modeling and references to earlier studies in  \citet{2021ApJ...917...71Z}).
{ In particular, fallback might be the key ingredient in explanation of properties of so-called central compact objects in SNRs  (see e.g., \citealt{1995ApJ...440L..77M, 2010RMxAA..46..309B, 2016MNRAS.462.3689I} and references therein).}

What is crucial for our study, in the case of fallback the NS is born as an Accretor. { Necessary conditions are easily fulfilled for total fallback mass $\gtrsim 10^{-5}\, M_\odot$ when $B\sim10^{13}$~G and $p_0\sim 20$~msec (see \citealt{2021ApJ...917...71Z} for detailed calculations and exact ranges of parameters).}

At early stages (hours to days), description of fallback can be quite complicated due to the action of a reverse shock, processes related to heating by radioactive elements, etc. \citep{1989ApJ...346..847C}. 
However, after some time (even tens of seconds in some models, see \citealt{2021arXiv210407493J}), fallback accretion rate eventually follows a simple power law:
\begin{equation}
    \dot{M} \sim t^{-5/3}.
\end{equation}
With such rapidly decaying rate of fallback accretion, most part of matter is accumulated within few minutes, see \cite{2021arXiv210407493J} for recent numerical calculations of fallback for $t<10^5$~s.  

If the star is isolated, fallback can last quite long. 
However, here we are interested in the situation in a close binary system which might be very different. In course of explosion, expelled matter mostly leaves the Roche lobe of the future compact object. So, falling back it might feel joint gravity of both components, and the compact object is significantly less massive in the situation we are interested in.
In addition, outside the Roche lobe the matter interacts with the stellar wind of the secondary component (here we consider only systems with wind accretion from massive donors). 
All in all, we can assume that mostly matter which do not leave the Roche lobe of the compact object can successfully fall back onto it. 

We estimate the Roche lobe radius, $r_\mathrm{L}$ using the standard formula \citep{1983ApJ...268..368E}: 
\begin{equation}
    \frac{r_\mathrm{L}}{a} = \frac{0.49 q^{2/3}}{0.6 q^{2/3} + \ln{(1 + q^{1/3})}},
\end{equation}
where $q$ is mass ratio of components.
For the LXP~4.4, the mass of the companion star is $\sim 10 M_{\odot}$
and $ {r_\mathrm{L}}/{a} = 0.28$, where $a$ is the semi-major axis of the binary. In our estimates we use $a = 10^{12}$ cm.

The maximum time of fallback accretion in a binary can be estimated as a free-fall time from the boundary of the Roche lobe:
\begin{equation}
    t_{\text{ff}} = \upi \sqrt{\frac{r_L^3}{G M_x}} \approx 8 \text{ } \bigg(\frac{r_L}{2.5  \times 10^{11} \text{ cm}}\bigg)^{3/2} \text{ h }.
\end{equation}
Thus, intensive fallback accretion cannot last long in binary systems, and so we can neglect this stage in our modeling, assuming as initial values of parameters those that they obtain after the fallback is over. 

After a short period of fallback accretion, the interaction of the NS and the companion star will be determined by the intensity of the stellar wind. 
Thus, the accretion rate would rapidly decrease.
Correspondingly, the size of the magnetosphere changes dramatically approaching the light cylinder. At this moment the phase transition is expected.
The key point of our model is that the NS will come to this moment starting, { at birth,} from the stage of accretion { due to the fallback}. This leads to the necessity to account for the hysteresis effect. 
Such scenario was not considered in other works on this topic.

\subsection{Algorithm of calculations}

We start our calculations when a brief period of fallback is over, and the external medium is determined by wind of the donor.
At this moment the NS has a period $p_0$, magnetic field $B_0$, and the accretion rate (from the wind) is $\dot{M}$.

With a fixed $\dot{M}$ for different combinations of  $p_0$ and $B_0$ one of three  situations may occur at the end of fallback:\\
1. If $R_\mathrm{m} < R_\mathrm{c}$ then the NS remains an Accretor;\\
2. If $R_\mathrm{c} < R_\mathrm{m} < R_\mathrm{l}$ --- the NS becomes a Propeller;\\
3. If $R_\mathrm{m} > R_\mathrm{l}$ --- the NS becomes an Ejector.

If the NS does not become an Ejector at this moment, it will not enter this stage in the future (for non-decreasing rate of capture of external matter) and will start accreting quite fast if it is a Propeller.

Further, corresponding to the current stage  calculations of the evolution of the period are carried out accordingly to eq.~(\ref{magrot}) for Ejector,  eq.~(\ref{eq:p_dot_prop}) for Propeller, and eq.~(\ref{eq:p_dot_acc}) for Accretor. Calculations are performed numerically using the SciPy library \citep{scipy}.

For each stage, we determine conditions at which
this stage is over. 
For Ejector this condition is:
\begin{equation}
    R_\mathrm{stop} = \max(R_\mathrm{l}, R_\mathrm{G}).
\end{equation}
Here $ R_\mathrm{stop} = R_\mathrm{Sh}. $
Usually, $R_\mathrm{l} < R_\mathrm{G}$, so this condition is satisfied when $R_\mathrm{Sh} = R_\mathrm{G}$.

For Propeller there are two conditions, as transition can happen either to the stage of accretion, or to the Ejector stage. The first one determines transition to an Accretor:
\begin{equation}
    R_\mathrm{stop} = \min(R_\mathrm{c}, R_\mathrm{G})
    \label{eq:event_P_A}
\end{equation}
here $ R_\mathrm{stop} = R_\mathrm{m}. $

The second determines transition to an Ejector:
\begin{equation}
    p = p_\mathrm{P \rightarrow E}.
\end{equation}

The condition for Accretor is the same as the condition of the Propeller $\rightarrow$ Accretor transition given by eq.~(\ref{eq:event_P_A}).

If one of such conditions is fulfilled then calculations are stopped, and we again check on which stage the NS is now. 
This procedure is organized as a cycle with an adjusted time step.
Calculations continue for sufficiently long time determined by life duration of the donor star (in some cases for illustrative purposes we run the code for a longer time to demonstrate all possible transitions under constant stellar wind rate).

\begin{figure}
	\includegraphics[width=\columnwidth]{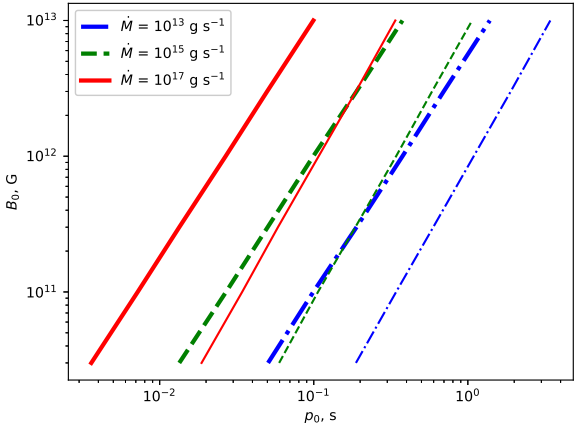}
    \caption{For three characteristic values of $\dot{M}$ we calculate critical values of $p_0$ and $B_0$. For a given $\dot{M}$, if the point is located to the left of the line, then a star is born as an Ejector. If initial period is higher or magnetic field is lower than critical values, then an NS will not be in ejector phase.
Thick and thin lines corresponds to calculations with presence and absence of the hysteresis effect, correspondingly.}
    \label{fig:P0_B0}
\end{figure}

\section{Results}

For three characteristic values of $\dot{M}$ we calculate critical values of $p_0$ and $B_0$ (Fig.~\ref{fig:P0_B0}). For a given $\dot{M}$, if the point is located to the left of the line, then a star is ``born'' (after a short episode of initial fallback accretion) as an Ejector. If initial period is higher or magnetic field is lower than critical values, then an NS will not be in the ejector phase and will start to accrete quite fast.
Bold and thin lines represent the presence and absence of the hysteresis effect, correspondingly.
Obviously, this effects plays a crucial role in avoiding the Ejector stage. 
The larger the accretion rate of the stellar wind, the easier it is to avoid the phase of ejection.

Equation of the line with hysteresis derived from the condition $p_0 = p_\mathrm{{P \rightarrow E}}$ has the following form:
\begin{equation}
    B_0 = \Bigg(\bigg(\frac{c p_0}{\upi}\bigg)^{7/2} \frac{2 \dot M \sqrt{2 G M_\mathrm{x}}}{R_{\mathrm{x}}^6}\Bigg)^{1/2}.
\end{equation}

Equation of the line without hysteresis is derived from the condition $p_0 = p_\mathrm{{E \rightarrow P}}$ can be written as:
\begin{equation}
    B_0 = \Bigg(\bigg(\frac{c p_0}{2 \upi}\bigg)^4 \frac{3 v \dot M}{2 
    R_{\mathrm{x}}^6}\Bigg)^{1/2}.
\end{equation}

\begin{figure}
	\includegraphics[width=\columnwidth]{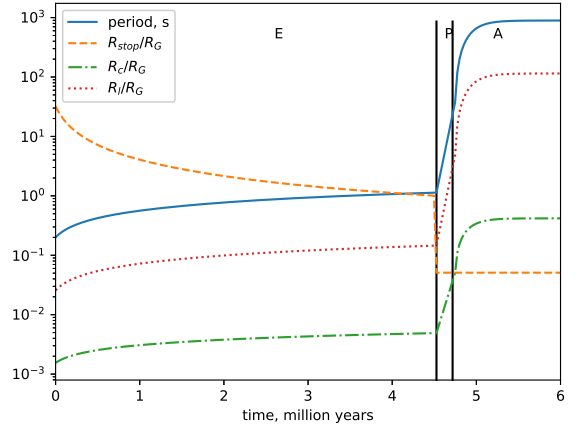}
    \caption{Evolution of the period and critical radii for an NS with $p_0 = 0.2$~s, $B_0 = 3 \times 10^{12}$G, $v=10^8$~cm~s$^{-1}$, $\dot M=7\times10^{13}$~g~s$^{-1}$, $P_\mathrm{orb} = 2.2$~d. The Ejector stage lasts long. Letters ``E'', ``P'', and ``A'' denote stages: Ejector, Propeller, Accretor, correspondingly. Solid vertical lines separate stages. Vertical axis is graduated to show spin period in seconds and different radii relative to $R_\mathrm{G}$.}
    \label{fig:standard}
\end{figure}

Fig.~\ref{fig:standard} shows a standard period evolution with the Ejector stage. Here we used parameters similar to those estimated or expected for LXP~4.4: $p_0 = 0.2$~s, $B_0 = 3 \times 10^{12}$~G, $v = 10^8$~cm s$^{-1}$, $\dot M = 7 \times 10^{13}$~g~s$^{-1}$, $P_\mathrm{orb} = 2.2$~d. 


\begin{figure}
	\includegraphics[width=\columnwidth]{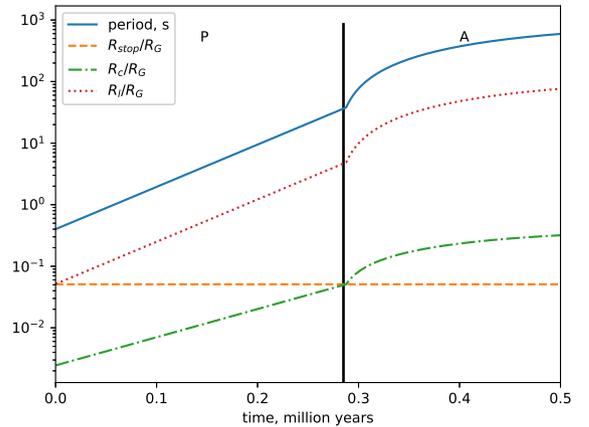}
    \caption{Evolution of the period and critical radii for an NS with $p_0 = 0.4$~s, $B_0 = 3 \times 10^{12}$~G, $v = 10^8$~cm~s$^{-1}$, $\dot M = 7\times 10^{13}$~g~s$^{-1}$, $P_\mathrm{orb} = 2.2$~d. Due to the hysteresis effect, it is possible to avoid the ejector stage and significantly reduce the time before the accretion.}
    \label{fig:hyst}
\end{figure}

With a slightly larger initial period ($p_0 = 0.4$~s, all other parameters are the same) it is possible to avoid the Ejector stage. In this situation, the NS starts to accrete much faster (Fig.~\ref{fig:hyst}).
Notice that $p_0 = 0.4$~s is a reasonable value (see e.g., \citealt{2013MNRAS.432..967I}), not as large as the initial period required for the condition that the NS is always at the Accretor stage.


\begin{figure}
	\includegraphics[width=\columnwidth]{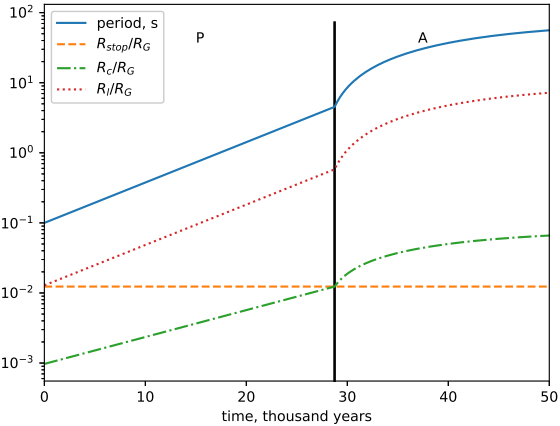}
    \caption{Evolution of the period and critical radii for an NS with $p_0 = 0.1$~s, $B_0 = 3 \times 10^{12}$~G, $v = 10^8$~cm~s$^{-1}$, $\dot M = 10^{16}$~g~s$^{-1}$, $P_\mathrm{orb} = 2.2$~d. With a higher accretion rate it is possible to start accretion in less than $10^5$ years.}
    \label{fig:M_dot_16}
\end{figure}

With larger $\dot M$, corresponding to luminosities $\sim 10^{36}$~erg~s$^{-1}$ (see Table 1), we can significantly shorten the time necessary to come to the phase of accretion. A corresponding example is given in Fig.~\ref{fig:M_dot_16}. Thus, it is possible to find a realistic set of parameters ($p_0, B_0, \dot M$) for which accretion starts before a SNR is dissipated. 

\section{Discussion and conclusions}

In this paper we presented a toy model of early evolution of an NS in a close binary system with a massive companion, accounting for fallback and peculiarities of Propeller~$\rightarrow$~Ejector transition. Below we comment on several aspects of this problem, some of which will be subject of future detailed consideration.

In many discussion of NS evolution different authors make diverse assumptions which can result in distinct conclusions. E.g., 
 \citet{2020MNRAS.494...44H} used $R_\mathrm{m} = R_\mathrm{l}$ as the condition of the E~$\rightarrow$~P transition. On our opinion, this is an oversimplification (see e.g., \citealt{1992ans..book.....L}). Instead, $R_\mathrm{Sh} = R_\mathrm{G}$ condition should be used, which makes the duration of the ejector phase significantly longer making LXP~4.4 even more puzzling in the standard framework without the hysteresis effect.  

Another typical assumption, already noticed above, is related to the equilibrium period, $p_\mathrm{eq}$. Often it is assumed that it can be defined by condition $R_\mathrm{A}=R_\mathrm{c}$. We denote such period as $p_\mathrm{A}$. However, equilibrium is determined by the balance between spin-up and spin-down. For wind-accreting systems with low $\dot M$ there is a significant difference between $p_\mathrm{A}$ and $p_\mathrm{eq}$ which might be taken into account, e.g. in estimates of the magnetic field. 

Equation for the equilibrium period might be different for low luminosities in the regime of so-called settling accretion \citep{2011evhe.confE..17P,2012MNRAS.420..216S}. This would change the estimate of the magnetic field:
\begin{equation}
B\approx 0.24\times 10^{12} \, \left(\frac{(p/100\ \mathrm{s})}{(P_\mathrm{orb}/100\, \mathrm{d})} \right)^{11/12} \dot M_{16}^{1/3} v_8^{-11/3}\, \mathrm{G}.
\end{equation}
For example, for parameters of LXP~4.4 we obtain: $B\approx 8.6 \times 10^{10} \, v_8^{-11/3} \, \mathrm{G}$. 
With such low field, { which are not unexpected for significant fallback amount (see \citealt{1995ApJ...440L..77M, 2016MNRAS.462.3689I} and references therein), } an NS might spend a very long time before starting to accrete from a stellar wind (however, notice strong dependence on the wind velocity). 
Significant initial fallback can allow accretion. However, details of such scenario { with the settling accretion} are out of the scope of the present study, and we plan to study it in future, in particular in application to isolated NSs.

\begin{figure}
	\includegraphics[width=\columnwidth]{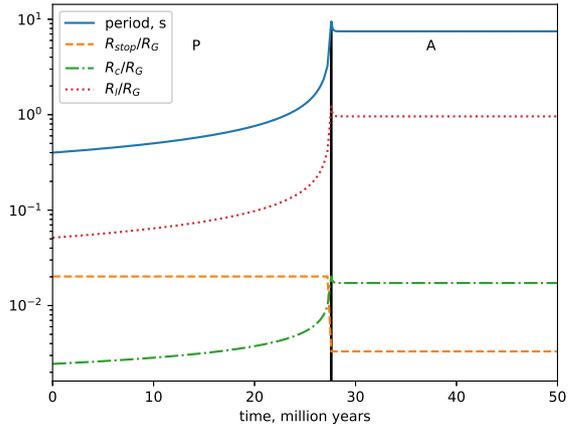}
    \caption{Evolution of the period and critical radii for an NS with $p_0 = 0.4$~s, $B_0 = 3 \times 10^{11}$~G, $v = 10^8$~cm~s$^{-1}$, $\dot M = 10^{16}$~g~s$^{-1}$, $P_\mathrm{orb} = 2.2$~d. Here we use equation for the spin-down rate at Propeller from Davies \& Pringle (1981). }
    \label{fig:D_P}
\end{figure}

There are many uncertainties in  magneto-rotational evolution of NSs.
Perhaps the most incomprehensible is the Propeller stage. { In particular, the spin-down rate at this phase of evolution which determines duration of the Propeller stage is uncertain. 
Several different formulas have been proposed to describe period evolution at this stage (see e.g., \citealt{1995AZh....72..711L}).

One of popular descriptions of the spin-down torque at the Propeller stage is the following equation \cite{1981MNRAS.196..209D}:}
\begin{equation}
    K_\mathrm{sd}^\mathrm{(P)}= \left( \frac{R_\mathrm{c}}{R_\mathrm{m}}\right)^{3/2} \frac{\mu^2}{R_\mathrm{m}^3}.
\end{equation}
{ Here a different formula for $R_\mathrm{m}$ is used:}
\begin{equation}
    R_\mathrm{m} = \left(\frac{R_\mathrm{G}}{R_\mathrm{A}}\right)^{2/9} R_\mathrm{A}.
\end{equation}
{ For realistic parameters, usage of this equation makes duration of the Propeller stage longer. This is illustrated in Fig. \ref{fig:D_P}. }


In this paper we do not study influence of different variants of propeller spin-down on the time before the onset of accretion. Still, this might be analysed, especially if the task is to reproduce properties of specific systems.

We also do not take into account the potentially possible stage of a subsonic propeller. It may occur when $R_\mathrm{m} < R_\mathrm{c}$, but the accretion is ineffective compared to the Accretor stage because of high temperature and the fact that the sound speed is higher than the rotation speed of the magnetosphere (\citealt{1981MNRAS.196..209D}, \citealt{2003A&A...399.1147I}).
At the subsonic propeller stage, the spin-down rate is lower, than at the classical (supersonic) propeller \citep{2005A&AT...24...17P}. Existence of such stage might postpone the onset of accretion. It can be especially important at low accretion rates (e.g., in case of isolated NSs).

As it was noted above, rapid onset of accretion can be explained by relatively large $p_0$.
However, the initial spin period of the NS in LXP~4.4 is not expected to be long due to two reasons. At first, because of interaction in a very tight binary the hydrogen envelope of the progenitor might be lost. So, rotation of the core could avoid significant spin-down via interaction with long living expanded envelope (see e.g., discussion in \citealt{2016MNRAS.463.1642P}  and references therein). Then, after losing its envelope  the progenitor could experience tidal spin-up. Depending on the time scale, the core could even reach tidal synchronization with the orbital period.  

In our toy model we do not consider disc formation from fallback matter. However, this phenomenon might be quite typical \citep{2021arXiv210407493J}. If a disc is formed, then it might be even easier to avoid the Ejector stage and make a soft transition to the wind accretion, may be even without an intermediate propeller phase. As most part of the falling back matter is accreted within a short time after a bounce, disc formation might not be influenced by the secondary companion. 

Intensive fallback might influence magnetic field of a newborn NS \citep{1995ApJ...440L..77M}. Technically,  field decrease is accounted for in our model, as we take as $B_0$ the value after the intensive phase of the fallback. So, we expect that NSs avoiding the Ejector stage might have lower fields (see Fig.~1) due to the fallback. In particular, we have to note that so-called central compact objects in SNRs (see a brief recent review in the introduction of \citealt{2021A&A...651A..40M}) have parameters which allow them to avoid the Ejector stage in binaries similar to those shown in Table 1. 
However, we ignore modification of the field structure due to fallback (see e.g., \citealt{2016MNRAS.462.3689I}) and field re-emergence \citep{2011MNRAS.414.2567H}. We hope to analyse these effects in future studies, as well as possible evolution of the angle $\chi$ in eq.~(\ref{magrot}), as the latter also can be an important ingredient of HMXBs evolution (see e.g. a recent analysis in \citealt{2021MNRAS.505.1775B} and references therein).

Hysteresis might also influence evolution of isolated NSs (and those in wide binaries which do not experience significant influence of the secondary component). Due to a long fallback stage an NS can spin-down significantly during the propeller phase, thus avoiding becoming a radio pulsar. This might influence fate of old isolated NSs allowing them to come the stage of accretion within the Galactic lifetime even for standard and, may be, lower magnetic fields in contrast to the standard scenario (\citealt{2010MNRAS.407.1090B}).
This will be the subject of our future studies.

Our final conclusions are as follows. 
 Presence of the fallback stage can significantly modify the evolution and observational properties of neutron stars in binary systems.
We demonstrate that the hysteresis effect, proposed by Shvartsman, allows a young neutron star in an HMXB to avoid the Ejector stage in the case of preceding fallback episode. With  magnetic fields $\sim 10^{12}$~G and initial spin periods $\sim 0.1$~--~0.2~s NSs can avoid the Ejector stage if the accretion rate is $\gtrsim 10^{14}$~--~$10^{15}$~g~s$^{-1}$. 
Such NSs can start accreting in a relatively short period, sometimes shorter than the lifetime of a SNR.
Some of  HMXBs in SNRs can be examples of such systems. However, detailed robust modeling of particular systems might need more sophisticated calculations.


\section*{Acknowledgements}
We thank drs. Andrei Igoshev and Anton Biryukov for comments on the manuscript and the anonymous referee for useful comments. 
This work was supported by the Russian Science Foundation, grant 21-12-00141.  

\section*{Data Availability}

 Observational data used in this paper are quoted from the
cited works. Data generated from computations are reported
in the body of the paper. Additional data can be made available upon reasonable request.




\bibliographystyle{mnras}
\bibliography{example} 








\bsp	
\label{lastpage}
\end{document}